\documentclass[aps,pre,floatfix,nofootinbib,showpacs,twocolumn]{revtex4}
\usepackage{amsmath}    
\usepackage{graphicx}
\usepackage{amssymb}
\begin{document}
\author{George Thomas\footnote{electronic address: george@iisermohali.ac.in}  
and Ramandeep S. Johal\footnote{electronic address: rsjohal@iisermohali.ac.in}}
\affiliation{Department of Physical Sciences, Indian Institute of Science Education and Research Mohali,\\
 Knowledge city, Sector 81, P.O. Manauli 140306, Ajit Garh, India.}
\title{Friction due to inhomogeneous driving of coupled spins in a quantum heat engine}
\begin{abstract}
We consider two spin-1/2 particles with isotropic Heisenberg interaction,
as the  working substance of a quantum  heat engine.  
We observe a frictional effect on the adiabatic branches of the heat cycle, 
which arises due to an inhomogeneous driving at a finite rate 
of the external  magnetic field. The frictional effect is characterized by  
entropy production in the system and reduction in the work extracted. 
 Corresponding to a sudden and a very slow driving, we find expressions for the lower and upper 
 bounds of work that can be extracted on
 the adiabatic branches. These bounds are also confirmed with numerical simulations
 of the corresponding Liouville-von Neumann equation. 
 
\end{abstract}
\maketitle
\section{Introduction}
The interplay between thermodynamic properties and quantum behavior 
has been studied through several models of quantum heat engines 
\cite{Scovil, Alicki1979,Geva1992, Kosloff2002, Kosloff2003, Scully, Mahler2004, Kieu2004, 
Kosloff2006, TZhang, AJM2008, Lutz, Noah2010, GJ2011, Abe2011, Kosloff2012, Huang2013}.
Even though quantum systems show non-classical features like entanglement, coherence and so on,
  these models are often found to be consistent with thermodynamic interpretations \cite{Scully,GJ2011,Leff}.
On the other hand, some models of these thermal machines have been reported to 
show an unexpected behavior such as extraction of work from a single heat bath \cite{Scully}, 
 exceeding Carnot efficiency \cite{Lutz} and cooling to absolute zero \cite{Kolar2012}.
In this paper, we focus on the interesting phenomenon  of intrinsic friction in  quantum engines 
\cite{Kosloff2002, Kosloff2003, Kosloffarxiv, Rezek2010}. This effect arises due to 
non-commutativity of the internal and the external part of the Hamiltonian
leading to non-commutativity of the Hamiltonians at different times.
 Further to reduce friction, an effect called quantum lubrication
has also been proposed \cite{Kosloff2006}. 
In order to better understand intrinsic friction and its relevance for analysis of dissipation
in quantum systems, it seems interesting to look for this effect in other 
similar models.
 
In the quantum heat engine that we discuss below, the working medium (system) consists of  two 
spin-half particles with Heisenberg interaction, kept in an external magnetic field.
The system is driven by selectively changing the external field in finite time, such that 
the field on either spin is different (inhomogenous). In this case, we observe the frictional
effect. But if the field on both spins remains homogeneous, then friction is absent.
As expected, we find that if the driving that creates inhomogeneity of fields on the spins, is performed
very slowly, the friction effect is again absent. 
 An analogous system under an 
inhomogeneous magnetic field plays important role in quantum computing \cite{sarma2001}.
Thermal entanglement of such spin system \cite{Asoudeh2005} and similar models
\cite{Albayrak2009} have  also been studied.

We consider a heat cycle analogous to Otto cycle,
consisting of two adiabatic branches and two thermalization branches. On the 
thermalization  branches, the magnetic fields on both spins are kept constant. 
To simplify our model, we consider 
the latter branches to take sufficiently long time so that 
the system attains equilibrium with the bath at the end of the process. 
 Then the system is decoupled from the bath and a thermodynamically adiabatic
 process is carried out on the system. So the initial state 
of the system before the adiabatic process is a thermal state,
diagonal in the eigenbasis of the Hamiltonian. This will help us
to understand the coherence developed in the system during 
the adiabatic process \cite{Kosloff2003}. 
The development of coherence leads to an increase in the entropy of the system which is 
a signature of the friction observed in our model.

The paper is organised as follows. In section II, we introduce the model of the quantum heat engine.  
Entropy production in the adiabatic branches of the cycle is discussed in section III. 
 Section IV is devoted for understanding the work extraction in our model. Here we discuss lower and 
upper bound of the work that can be extracted. 
Section V is devoted to discussion. We analyse the cycle using numerical 
simulations by alloting finite time 
to the adiabatic branches and conclude with a summary and future directions. 

\section{Model}
We consider two spin-half particles with isotropic exchange 
interaction, as the working substance for a 
quantum Otto cycle.
 In general, the Hamiltonian is written as $H=H_{\rm{int}}+H_{\rm{ext}}$, where
$H_{\rm ext}$ is the external Hamiltonian which can be
controlled and $H_{\rm{int}}$ is the internal Hamiltonian. 
In our model, we control in time, the magnetic field applied
to particle labeled 2. So  we have 
\begin{equation}
H_{\rm int} = J({\sigma^{(1)}}.{\sigma}^{(2)} + {\sigma}^{(2)}.{\sigma}^{(1)}), 
\end{equation}
\begin{equation}
H_{\rm ext}=B_1 \sigma^{(1)}_z+ B_2(t)\sigma^{(2)}_z,
\label{H}
\end{equation}
where  $\sigma^{(i)}=(\sigma_{x}^{(i)} ,\sigma_{y}^{(i)},\sigma_{z}^{(i)})$
are the Pauli matrices, $J$ is the isotropic exchange constant and $B_1$, $B_2(t)$ are the
magnetic fields applied along $z$-axis to the first 
and the second spin respectively.
So the magnetic field applied to the
individual spins are not always equal during the adiabatic branch which results in $[H_{\rm ext},H_{\rm int}]\neq0$.
This non-commutativity of the external and the internal Hamiltonian when leading to non-commutativity of the Hamiltonian 
at different times,
is the cause of internal friction in our model \cite{Kosloff2002, Kosloff2003}.

As a special case, we show in Section V that the non-commutative property of external and the internal Hamiltonian
by itself is not a sufficent condition for friction.

Now we analyse the system with inhomogeneous magnetic field in more detail. 
In this case the eigenbasis of the 
Hamiltonian is $\{|\psi_i\rangle; i=1,..,4\} \equiv$ \{$|\psi_1\rangle$, $|00\rangle$,  $|\psi_3\rangle$, $|11\rangle$\},
 where 
$|\psi_1\rangle$ and  $|\psi_3\rangle$ are  given by $b|10\rangle-a|01\rangle$ and $a|10\rangle+b|01\rangle$
 respectively and \{$|00\rangle$, $|10\rangle$,  $|01\rangle$, $|11\rangle$\} forms the computational basis. 
 Here $a=(y+\sqrt{1+y^2})/N$
and $b=1/N$,  where $N=\sqrt{1+(y+\sqrt{1+y^2})^2}$ and $y=(B_1-B_2(t))/4J$.
 The corresponding eigenvalues are  \{$-2J-K$, $2J-B_1-B_2(t)$,  $-2J+K$, $2J+B_1+B_2(t)$\},
 where  $K=4J(\sqrt{1+y^2})$. 
The equilibrium density matrix when the system is attached to a bath at temperature $T_e$, is given by
$\rho= \exp{(-H/T_e)}/Z$, where $Z={\rm Tr} (\exp{(-H/T_e)})$ is partition function of the system,
%
and we have set Boltzmann's constant to unity. The eigenvalues of $\rho$, or the occupation probabilities
of the energy levels, are given by
\begin{eqnarray}
P_1&=&e^{-(-2J-K)/T_e}/Z,\nonumber \\
P_2&=&e^{-(2J-B_1-B_2(t))/T_e}/Z, \nonumber \\
P_3&=&e^{-(-2J+K)/T_e}/Z,\nonumber \\
P_4&=&e^{-(2J+B_1+B_2(t))/T_e}/Z.
\label{probability}
\end{eqnarray}
Now we are ready to discuss the quantum heat cycle, which consists of the following four stages:

\textit{Stage 1}: The coupled-spins system is attached to a cold bath with temperature $T_1$.
 The system attains equilibrium with the bath. The magnetic field applied to the
first and second spins are identical ($B_1=B_2$). 
The density matrix  is
 diagonal in the Hamiltonian's eigenbasis. Because of the homogeneous magnetic field, 
the eigenstates $|\psi_1\rangle$ and  $|\psi_3\rangle$ are maximally entangled Bell
states, with $a=b=1/\sqrt{2}$. 
The occupation probability $\{p_j\}$ for the
 state with energy eigenvalue $\{E_j\}$ is calculated from Eq. (\ref{probability}) by setting $T_e=T_1$
and $B_2=B_1$. So the mean energy at the end of the first stage is Tr$(\rho H)=\sum_jE_jp_j$.

\textit{Stage 2}: In this stage, the system is isolated from the bath and it can
exchange only work with the surroundings.
 The magnetic field applied to the second spin is changed from $B_2(0)=B_1$ to $ B_2(t)=B_3$ 
in finite time  and the system may undergo a non-adiabatic evolution. By non-adiabatic evolution,
 we mean that the system may be driven fast enough so that the quantum adiabatic
theorem does not hold \cite{Fock,Kato}. 
The density matrix undergoes a unitary evolution. The eigenstates of  $H(t)$ are also
time dependent. In general, the eigenstates of $\rho(t)$ 
are not the same as $H(t)$.
In the infinitely slow limit ($t\rightarrow \infty$),
 the adiabatic theorem holds and eigenstates of the density matrix are identical
to the eigenvectors of the instantaneous Hamiltonian.

\begin{figure}[ht]
\begin{center}
\includegraphics{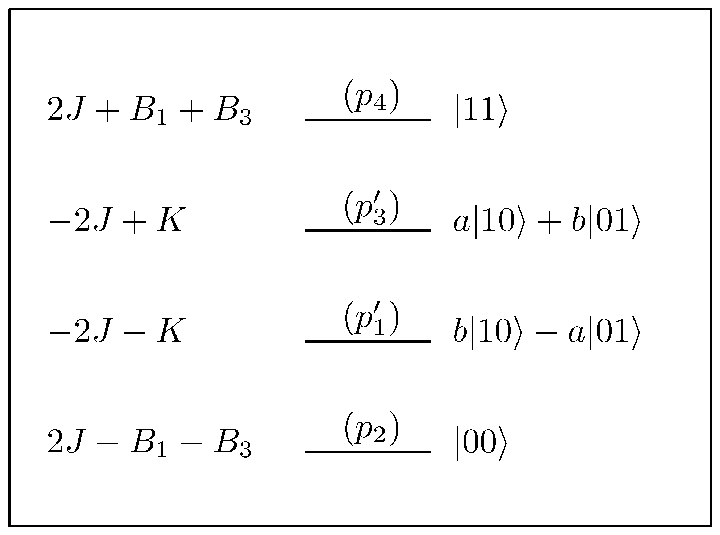}
\end{center}

\caption{A pictorial representation of eigenvalues and eigenstates of the
Hamiltonian at the end of first adiabatic process (stage 2).
The $\{ p_i' \}$ represent the populations in the energy eigenbasis $\{ |\psi_i' \rangle \}$.
 In the infinite time limit, we get $p_i'=p_i$ and the
eigenstates of the density matrix are same as that  of the Hamiltonian.}

\label{A1}
\end{figure}

So in case of fast driving, the final state of the system may not be diagonal in the eigenbasis of the final Hamiltonian.
When we project the final density matrix onto the eigenbasis of the Hamiltonian, the corresponding
occupation probability of the eigenstate of the Hamiltonian with eigenvalue $E_j'$
  is given as $p_j'={\rm Tr}\left(|j\rangle\langle j| \rho(t)\right)$, where $|j\rangle$ is the eigenvector of the final Hamiltonian.
 A pictorial representation is shown in Fig. \ref{A1}. At the end of the second stage, the mean energy can be written as
Tr$(\rho(t) H(t))= \sum_jE_j'p_j'$. The difference of the initial and the final mean energy 
during the adiabatic process is equal to the work performed: 
$W_I=\sum_j E_j p_j-\sum_jE_j' p_j'$.

\textit{Stage 3}: The system under inhomogeneous magnetic field is attached
 to a hot bath with temperature $T_2$ and it attains equilibrium by absorbing heat from the bath.
The occupation probabilities ($q_j$) are calculated from Eq. (\ref{probability}) by putting $B_2=B_3$ and $T_e=T_2$.
At the end of the third stage, the system is in a thermal state with mean energy $\sum_jE_j'q_j$.

\textit{Stage 4}: The system again undergoes a unitary 
evolution by a change of the magnetic field of the second spin from $B_3$ to $B_1$,
whereby the energy levels change from $E_j'$  back to $E_j$.
 The occupation probabilities $q_j'$ in the eigenstates of the 
Hamiltonian are calculated by projecting the density matrix onto the eigenbasis of the Hamiltonian.
So the  mean energy at the end of the process is $\sum_jE_jq_j'$. The difference in the mean energy due to this process is
$W_{II}=\sum_jE_j'q_j-\sum_jE_jq_j'$.

 To close the cycle, the system is again brought
in contact with cold bath. The system releases on average  an amount of heat to cold bath.
As we show below, $W_I$ and $W_{II}$ are the work done {\it by} and {\it on} 
the system, respectively. 

\section{Dynamics on adiabatic branch and entropy}
Now we analyse the irreversibility associated with the adiabatic branch by
 quantifying the entropy production. 
The adiabatic process is represented by a unitary process so that after time $t$, 
the system-state evolves to $\rho(t)=U(t,0)\rho(0)U^{\dagger}(t,0)$, where 
$U={\cal T} \exp{(-i\int_0^tH(t')dt')}$.
 The von Neumann entropy $S_v$  remains constant throughout the process. 
 But  energy-entropy $S_e$, defined  with the occupational probabilities of the energy levels, changes. 
 $S_e$ in the initial state is given by
$ -\sum_ip_i\ln p_i$, where $p_i={\rm Tr}(|\psi_i\rangle \langle \psi_i|\rho(0))$. 
Since the initial state is a thermal state, we have $S_e=S_v$.
But after the finite-time adiabatic step, $S_e$  increases where as $S_v$ remains unchanged.
Initially, we have $[H(0),\rho(0)]=0$.
Two of the  eigenvectors  $| 00 \rangle$ and $|11\rangle$ of the Hamiltonian are not 
functions of the applied magnetic field and hence are independent of time. 
So if the system is in any of these ${\it two}$ eigenstates, it
 will remain there during the process. 
Thus the initial population in theses states remains
constant throughout the adiabatic process. 

But the  eigenvectors 
 $|\psi_1 \rangle$ and $|\psi_3\rangle$ of hamiltonian depend on  the magnetic field and hence are time dependent. 
 So if the system is initially in one of these states, then changing the
 Hamiltonian with a finite 
rate results in a non-adiabatic evolution. In 
in other words, the final state of the system is then not an eigenstate  of the
final Hamiltonian. Let  the eigenvectors of the final Hamiltonian be given as
 $\{ |\psi_1'\rangle$, $|00\rangle$, $|\psi_3'\rangle$, $|11\rangle \}$ and 
the set of the eigenvectors of the final density matrix
is \{$|\phi_1'\rangle$, $|00\rangle$, $|\phi_3'\rangle$, $|11\rangle$\}. 
Since  $|\phi_1'\rangle$ and $|\phi_3'\rangle$ are
orthogonal to each other as well as to $|00\rangle$  and  $|11\rangle$, 
we can express the kets $|\phi_1'\rangle$ and $|\phi_3'\rangle$ as  linear combinations
of  $|\psi_1'\rangle$ and $|\psi_3'\rangle$ as
\begin{eqnarray}
|\phi_1'\rangle&=&\cos(\delta/2)|\psi_1'\rangle + \sin(\delta/2)|\psi_3'\rangle, \nonumber \\
|\phi_3'\rangle&=&\sin(\delta/2)|\psi_1'\rangle - \cos(\delta/2)|\psi_3'\rangle,
\label{phipsi}
\end{eqnarray}
where $0 \le \delta \le \pi$. 
Now consider a projection of the system-state on the eigenbasis of Hamiltonian.
 Two of the populations remain unchanged such that  
$p_2'=p_2$ and $p_4'=p_4$. The occupation probabilities 
for the eigenstates $|\phi'_1\rangle$ and $|\phi'_3\rangle$
are $p_1$ and $p_3$ respectively. Now project the density matrix onto the eigenbasis 
$\{ |\psi_i' \rangle \}$ of the final Hamiltonian. From Eq. (\ref{phipsi}),
we get the occupation probabilities corresponding to $|\psi'_1\rangle$ and $|\psi'_3\rangle$ 
as
\begin{eqnarray}
p_1'&=&p_1\cos^2(\delta/2)+p_3 \sin^2(\delta/2), \nonumber \\
p_3'&=&p_1\sin^2(\delta/2)+p_3 \cos^2(\delta/2).
\end{eqnarray}
Due to $p_1 > p_3$, we can write 
\begin{eqnarray}
p_1&\geq&p_1'\geq p_3, \nonumber \\
p_1&\geq&p_3'\geq p_3.
\label{p1p3}
\end{eqnarray}
As the difference between $p_1'$ and $p_3'$ gets reduced
as compared to the one between $p_1$ and $p_3$, and recalling
that $p_2'=p_2$ and $p_4'=p_4$, the distribution 
$\{ p_i' \}$ is more uniform than $\{ p_i \}$, we have
\begin{equation}
-\sum_ip_i'\ln p_i'\geq -\sum_i p_i\ln p_i,
\label{entropy}
\end{equation}
which signifies that the energy-entropy $S_e$ increases in the finite-time adiabatic process.
In the infinite time process ($t \rightarrow \infty$), the system undergoes
quantum adiabatic evolution and in this limit $S_e$ remains unchanged. 
The  total entropy production versus the total time allocated to
adiabatic branch will be discussed in Section V.
%
\section{Work}
 The work is performed by or on the system only during the adiabatic branches 
 i.e. in stages 2 and 4, when the evolution of the system is governed by 
 Liouville-von Neumann equation (with $\hbar=1$)
\begin{equation}
 \frac{d\rho(t)}{dt}=-i\left[H(t),\rho(t)\right].
\label{Liouville}
\end{equation}
The instantaneous mean energy of the system is given by ${\rm Tr}(\rho(t)H(t))$.
Differentiating with respect to time we get
\begin{equation}
 {\rm Tr}\left(\frac{d(H(t)\rho(t))}{dt}\right)= {\rm Tr}\left(H(t)
\frac{d\rho(t)}{dt}\right)+{\rm Tr}\left(\frac{dH(t)}{dt}\rho(t)\right),
\label{dmeanE}
\end{equation}
In general, comparing with the first law of thermodynamics, we identify 
 \cite{Alicki1979} the first term on the right hand side as the
 rate of heat flow ($\dot{Q}$) and the second term as the  power ($\wp$).

For an adiabatic process, the first term above on the right hand side
vanishes due to Eq. (\ref{Liouville}).

Upon integrating the power, we get the expression for work as
\begin{eqnarray}
W = \int_0 ^{t} \wp dt &=& \int_0^{t}{\rm Tr}\left(\frac{dH(t')}{dt'}\rho(t')\right)dt', \nonumber \\
&=&\int_0^{t}{\rm Tr}\left(\frac{d(H(t')\rho(t'))}{dt'}\right)dt'.
\label{workderivation}
\end{eqnarray}
Thus the work performed during the adiabatic process 
lasting for a time interval $t$, is equal to the
change in the mean energy of the system, upto time $t$.


Furthermore, it can be shown that the work done in a infinitely slow  process is always higher than
the work done in a finite-time process.
Thus the lower bound for work extracted is obtained for an extremely fast process ($ t\rightarrow 0$).
To evaluate the lower bound, we assume that the density matrix of the system remains unchanged.
In case of equilibrium with the cold bath, the initial density matrix is given as
\begin{equation}
 \rho = p_1\arrowvert\phi_1\rangle\langle\phi_1| + p_2\arrowvert00\rangle\langle00| + 
p_3\arrowvert\phi_3\rangle\langle\phi_3| + p_4\arrowvert11\rangle\langle11|,
\label{rho}
\end{equation}
where $|\phi_{1} \rangle = (\arrowvert10\rangle-\arrowvert01\rangle)/\sqrt{2}$ and
  $|\phi_{3} \rangle = (\arrowvert10\rangle+\arrowvert01\rangle)/\sqrt{2}$.
Since the system is in thermal state, the initial Hamiltonian commutes 
with the density matrix and both have the  same set of 
eigenvectors. In the sudden limit ($t\rightarrow0$), the density matrix
remains the same as the initial, because $U(0,0)=I$. But the Hamiltonian is changed to
\begin{eqnarray}
 H=&-&(2J+K)\arrowvert\psi'_1\rangle\langle\psi'_1|+(2J-B_1-B_3)\arrowvert00\rangle\langle00| \nonumber \\
&+&(-2J+K)\arrowvert\psi'_3\rangle\langle\psi'_3|+ (2J+B_1+B_3)\arrowvert11\rangle\langle11|, \nonumber \\
\end{eqnarray}
where $|\psi'_{1} \rangle =b\arrowvert10\rangle-a\arrowvert01\rangle$ and
  $|\psi'_{3} \rangle =a\arrowvert10\rangle+b\arrowvert01\rangle$.
 Now we find the population of the corresponding
eigenstates of the Hamiltonian by projecting the density matrix onto the eigenbasis of Hamiltonian as
\begin{eqnarray}
p_1'=\langle\psi'_1|\rho|\psi'_1\rangle=\frac{(p_1+p_3)}{2}-ab(p_3-p_1), \\
 p_3'=\langle\psi'_3|\rho|\psi'_3\rangle=\frac{(p_1+p_3)}{2}+ab(p_3-p_1),
\label{p3'p1'}
\end{eqnarray}
while $p_2'=p_2$  and $p_4'=p_4$. 
Similarly for the second adiabatic process where the Hamiltonian is returned
to its initial stage with eigenbasis $\{ | \psi_i \rangle \}$, we obtain
upon projecting the density matrix $\tilde{\rho}$ for this process, as
\begin{eqnarray}
q_1'=\langle\psi_1|\tilde{\rho}|\psi_1\rangle=\frac{(q_1+q_3)}{2}-ab(q_3-q_1), \\
 q_3'=\langle\psi_3|\tilde{\rho}|\psi_3\rangle=\frac{(q_1+q_3)}{2}+ab(q_3-q_1),  
\label{q3'q1'}
\end{eqnarray}
and $q_2'=q_2$  and $q_4'=q_4$.

Now the work extracted in complete cycle ($W=W_I + W_{II}$) with fast adiabatic processes is given by
\begin{equation}
 W^{\rm fast} =\sum_i p_i E_i - \sum_i p_i' E_i' + \sum_i q_iE_i' - \sum_i q_i'E_i. 
\label{Wfast}
\end{equation}
Using the probabilities calculated above for extremely fast (sudden) processes, we get the lower bound 
of work 
\begin{eqnarray}
 W_{\rm lb}=& & (B_3-B_1)(q_4-q_2+p_2-p_4) \nonumber \\
            && + (q_3-q_1)(K-8Jab). 
\label{wlb}            
\end{eqnarray}
The upper bound for work is obtained for the slow process ($t\rightarrow \infty$). 
According to quantum adiabatic theorem the system remains in the instantaneous eigenstate
of the Hamiltonian. The work expression is in general written as
\begin{equation}
 W^{\rm slow} = \sum_i p_i E_i - \sum_i p_i E_i' + \sum_i q_i E_i'  - \sum_i q_i E_i,
\label{Wslow}
\end{equation}
yields the upper bound for the extractable work,
\begin{eqnarray}
 W_{\rm ub} = && (B_3-B_1)(q_4 - q_2 + p_2 - p_4) \nonumber \\
&& +(q_3-q_1+p_1-p_3)(K-4J).
\label{wub}
\end{eqnarray}
These bounds are compared with the finite-time work in Fig. (\ref{figurework}).
\section{Discussion}
\begin{figure}[h]
\begin{center}
\includegraphics[width=7.8cm,height=5cm]{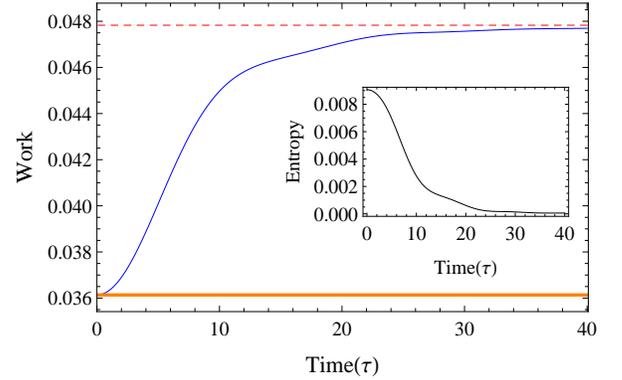}
\end{center}
\caption{Work obtained in a cycle versus the total time ($\tau$) allocated for both 
the adiabatic processes.  Here we use
 $B_1=B_2(0)=3$, $B_3=4$, $J=0.1$,
 $T_2=2$ and $T_1=1$. The  work is bounded above by $W_{\rm ub}$ (Eq. (\ref{wub}),
 dashed line). The thick horizontal line depicts the lower bound $W_{\rm lb}$,  Eq. (\ref{wlb}), 
obtained for a sudden adiabatic process ($\tau \to 0$).
The inset shows total entropy production on the adiabatic branches 
versus total time ($\tau$). As $\tau$ is increased, the total
entropy production reduces monotonically to zero and the frictional effect vanishes.}
\label{figurework}
\end{figure}

Analytic expressions for work  can be derived both in the case of a very slow driving and
a sudden one. 
To estimate the finite-time evolution of the system on the adiabatic branch, we have to
integrate the Liouville-von Neumann equation, Eq. (\ref{Liouville}).
We accomplish this using the fourth-order Runge-Kutta method \cite{Mahler2006}.
In the first adiabatic process, $B(t)$ changes from $B_2(0)$ to $B_3$. 
This is modeled by applying a pulse $B_2(t)=B_2(0)+ (B_3-B_2(0))\sin{({\pi t}/{\tau})}$
for a time $t=\tau/2$, where $\tau$ is half of the time period.
Similarly the second adiabatic process is 
done by applying a pulse $B_2(t) = B_3 + (B_2(0)-B_3)\sin{({\pi t}/{\tau})}$
for the same time interval. Thus we allot equal time intervals to both
the adiabatic branches.
The total  work performed and the total entropy production due to the 
finite time process are plotted in Fig. (\ref{figurework}). As discussed
in the previous section, the work extracted decreases monotonically with a finite rate of driving.
This is also reflected in the corresponding increase of entropy production
in the finite-time case.

Let us consider a cycle in which the magnetic fields applied 
 to the first and second spins in {stage 1} 
 have different values, $B_1(0)$ and $B_2(0)$ respectively.
 In this case the internal and external part of the hamiltonian
 do not commute with each other. Now suppose that 
 during the  first  adiabatic process, $B_1$ and $B_2$
 vary at equal rates so that the difference  ($\Delta B = B_1(t) - B_2 (t)$) 
keeps constant during the process. As we have seen in section II, the parameters   
$a$ and $b$ appearing in the eigenbasis of the Hamiltonian, are functions of $J$ and $\Delta B$. 
Since $\Delta B$ remains constant during  the adiabatic process,
 the energy eigenstates of the Hamiltonian become
time independent which implies that Hamiltonians at
different times commute with each other and so friction is
absent in this case.  Using the similar argument,
 no friction is expected
on the second adiabatic process, when the magnetic
fields are restored to their initial values ($B_1 (0)$ and $B_2(0)$). 
This serves as an example to appreciate that the non-commutative property of the internal and external
Hamiltonian caused by the inhomogeneous magnetic fields may not always
lead to non commutativity of Hamiltonian at different times to cause friction.
 Rather the inhomogeneous driving in which  $\Delta B$ changes with
 time leads to the non commutative property of the 
Hamiltonian at different times and thereby to frictional effect. 
 

To conclude, we have studied a model of quantum heat engine where the 
inhomogeneous driving at a finite rate, of the components of the quantum working medium
 leads to a frictional effect. This 
  effect  is characterized by  increase in the entropy of
 the system. As expected of a thermodynamic system, 
the entropy production leads to decrease in the work obtained from a cycle. 
The work is plotted versus the time allotted 
for the adiabatic branches. The amount of work that can be obtained from 
our model is bounded from both above and below. The upper bound
is obtained for a slow process where the frictional effect vanishes and
quantum adiabatic theorem holds, while the lower bound is obtained for a sudden process.
Some interesting future problems include the study of frictional 
effect on models with anisotropic interactions and with 
systems using higher number of spins.
The possibility of quantum lubrication \cite{Kosloff2006}
 to reduce the intrinsic friction can also be studied.
\section*{Acknowledgements}
GT acknowledges  financial support from IISER Mohali. RSJ
gratefully acknowledges reearch grant from the Department of Science and Technology, India,
under project No. SR/S2/CMP-0047/2010(G) titled,  
{\it Quantum heat engines: work, entropy and information at nanoscale}.
\end{document}